\documentclass[12pt,preprint]{aastex}
\newcommand{\etal}{{et al.} }

\newcommand{\chandra}{{\it Chandra} }

\newcommand{\hetg}{{\it HETGS} }

\newcommand{\fekalfa}{{Fe~K$\alpha$} }

\newcommand{\blue}[1]{{\bf }}

\shorttitle{CHANDRA HEG OBSERVATIONS Fe K$\alpha$ LINES IN AGN}

\begin{document}

\title{ 
On the X-ray Baldwin Effect in Active Galactic Nuclei Observed by the \chandra High Energy Grating } 

\author{X. W. Shu\altaffilmark{1}, J. X. Wang\altaffilmark{1}, T. Yaqoob\altaffilmark{2}, P. Jiang\altaffilmark{1}, Y. Y. Zhou\altaffilmark{1}}

\altaffiltext{1}{
CAS Key Laboratory for Research in Galaxies and Cosmology, 
Department of Astronomy, University of Science and Technology of China, 
Hefei, Anhui 230026, P. R. China, xwshu@mail.ustc.edu.cn, jxw@ustc.edu.cn}
\altaffiltext{2}{Department of Physics and Astronomy,
Johns Hopkins University, Baltimore, MD 21218, yaqoob@pha.jhu.edu}

\begin{abstract}
Using \chandra High Energy Grating (HEG) observations of 32 AGNs,  we present a systematic study of the X-ray Baldwin effect 
(XBE, i.e. the anti-correlation between narrow \fekalfa\ line EW and X-ray continuum luminosity for AGN samples) with the highest spectral resolution currently available. 
We have previously reported an anti-correlation with $EW~\propto~L_{\rm 2-10~keV}^{-0.22}$
in a HEG sample, and the correlation is much weaker after averaging multiple observations of individual AGNs ($EW~\propto~L^{-0.13}_{\rm 2-10~keV}$).
This indicates that rapid variation in
X-ray continuum plays an important role in producing the XBE, and such an 
effect should also be visible in individual AGNs.
In this paper, by normalizing the line EWs and continuum luminosities to the time-averaged values for each AGN in our sample with multiple HEG observations,  
we find a strong anti-correlation between EW and L$_X$
(EW/$\langle EW \rangle \propto (L/\langle L \rangle)^{-0.82\pm0.10}$), 
consistent with the XBE expected in an individual AGN if the narrow line flux remains constant while the continuum varies.  
This is first observational evidence that the  
\fekalfa line flux in a large sample of AGNs lacks of a corresponding response to the continuum variation, 
supporting that the narrow Fe-K line emission originates from 
a region far from the nucleus. 
We then performed Monte-Carlo simulations 
to address whether the global XBE can be produced by X-ray continuum variation solely, 
and found that such interpretation of XBE cannot be ruled out statistically.
One thus should be very cautious before reaching
any scientific conclusion based on an observed XBE.
\end{abstract}


\section{INTRODUCTION}
\label{intro}
The narrow (FWHM~$<10,000 \ \rm km \ s^{-1}$) 
 \fekalfa fluorescent emission at $\sim 6.4$~keV is a common feature in the X-ray spectra 
of active galactic nuclei (AGN)  
(e.g. Sulentic \etal 1998; Lubi\'{n}ski \& Zdziarski 2001;
Weaver, Gelbord, \& Yaqoob 2001; 
Perola \etal 2002; Yaqoob \& Padmanabhan 2004; 
Bianchi et al. 2007;  
 Winter \etal 2009; Shu, Yaqoob \& Wang 2010 (hereafter Paper I); Fukazawa et al. 2011). 
Such a narrow \fekalfa line is believed to be produced in cold, neutral 
 matter far from the nucleus (see Paper I, and references therein). 
Possible origins of the \fekalfa line include the outer regions of an accretion disk, 
the broad-line region (BLR), and a parsec-scale torus (Nandra 2006; Bianchi et al. 2008; 
Paper I; Liu \& Wang 2010; Shu, Yaqoob \& Wang 2011; Jiang, Wang \& Shu 2011).  
 In some cases, the line has been resolved by the \chandra high-energy grating (HEG; 
see Markert et al. 1995)   
with FWHM typically less than 5000 $\rm km \ s^{-1}$ (e.g. Kaspi et al. 2000; Yaqoob et al. 2001; 
Paper I; Shu et al. 2011). In Paper I, we measured the intrinsic width of the narrow \fekalfa line core in a large sample of type I AGNs using the \chandra HEG, and 
obtained a weighted mean of FWHM=$2060\pm230$ km s$^{-1}$. 
Interestingly, this width of the line is fully consistent with that obtained in 
type II AGNs observed with \chandra HEG (Shu et al. 2011), indicating a common origin 
of the narrow \fekalfa line in both AGN populations.  

The equivalent width (EW) of the \fekalfa line was found to be anti-correlated with 
the X-ray continuum luminosity (also known as X-ray Baldwin effect) in several AGN 
samples (e.g., Iwasawa \& Taniguchi 1993; Nandra et al. 1997b; Page et al. 2004; 
Jiang, Wang, \& Wang 2006; Bianchi et al. 2007; Winter et al. 2009; Chaudhary et al 
2010). Note the observed X-ray Baldwin effect in some samples could be (at least) partly due to the contamination of radio-loud objects (Jimen$\acute{\rm e}$z-Bail$\acute{\rm o}$n et al. 2005; Jiang et al. 2006). 
It is also interesting to note a recent work by Krumpe et al. (2010) who found that the X-ray Baldwin effect flattens above
$L_X\sim10^{44}$ erg s$^{-1}$, claiming that the anti-correlation (if exists) is only
relevant in radio-quiet AGNs with $L_X<10^{44}$ erg s$^{-1}$.

There are several physical factors that could produce the X-ray Baldwin effect.
One possibility is that  the covering factor 
and/or the column density of line-emitting material decreases with increasing X-ray continuum 
luminosity (e.g., Page et al. 2004). 
Alternatively, the X-ray Baldwin effect could be artificial due to the 
rapid X-ray continuum variation in AGNs.
Jiang et al. (2006) reported that X-ray continuum variation model can naturally
produce a global X-ray Baldwin effect in their composite Chandra/XMM sample.
In Paper I, with Chandra HEG data, we also found that averaging multiple observations of individual sources could significantly smear the X-ray Baldwin effect from
$EW~\propto~L^{-0.22}$ to $EW~\propto~L^{-0.13}$.
This clearly indicates that the observed X-ray Baldwin effect could at least 
partially attributed to continuum variation.


Another possibility is contamination from the Fe K$\alpha$ 
line emission from the accretion disk,  which
could be more and more ionized as the X-ray 
luminosity increases, leaving less low-ionization material to produce the Fe K$\alpha$ 
line at $\sim$ 6.4 keV (e.g., Nandra et al. 1997b; Guainazzi et al. 2006; Nayakshin 2000a, 2000b).
Therefore, it is essential to study the X-ray Baldwin effect for narrow Fe K line emission 
with the highest spectral resolution available, 
to isolate the narrow cores from the broad component.

Using a sample that consists only of HEG data\footnote{\chandra HEG affords best
spectral resolution currently available in the Fe K band, at 6.4 keV is $\sim$39 eV,
or $\sim$1860 km s$^{-1}$ FWHM.}, we can investigate the X-ray Baldwin
effect with a spectral resolution in the Fe K band that is nearly four times better
than in previous studies that relied on CCD data (or data with a resolution worse than CCDs), 
and therefore provide the best isolation of the narrow core
that is currently possible.
In this paper, we extend the results of Paper I to investigate  
the origin of the Baldwin effect, in particular taking into consideration of 
the effect of X-ray continuum variability. 
Throughout this paper, we adopt a cosmology of $\Omega_M=0.3, \Omega_{\lambda}=0.7$, 
and H$_0=70$ km s$^{-1}$ Mpc$^{-1}$.     
 
\section{OBSERVATIONS AND SPECTRAL FITTING}
\label{data}
Our study is based on data from 85 \chandra \hetg observations of 32 AGNs that are 
known to have 
X-ray absorbing column densities ($N_{\rm H}$) less than $5\times10^{22}$ cm$^{-2}$, 
as of 2010 Aug 1, filtering on several criteria as shown in Paper I . 
Note that heavily absorbed AGNs were excluded 
because their X-ray spectra are too complex, and  
the measurements of the narrow \fekalfa line and 
intrinsic continuum luminosities in such sources can become model dependent.    
Except for 15 new observations for NGC 4051 and Ark 564, all the data 
are taken from Paper I. 
We excluded 4 radio-loud objects (3C 120, 3C 273, 3C 382 and 4C 74.26) 
from our analysis, because their X-ray spectra 
might be contaminated by the relativistic jet. 
The \chandra data for the sample 
were reduced and HEG spectra were analyzed as described in Yaqoob \etal (2003).
Further details of the \chandra HEG data can be found
in Paper I.
Note that in our analysis, we fixed the emission-line width, $\sigma_{\rm FeK}$, 
at 1 eV (corresponding to $\sim$ 100 km s$^{-1}$ FWHM at 6.4 keV), a 
value well below the HEG resolution, in order to obtain a uniform measurements of 
the narrowest, unresolved core component of the \fekalfa 
line for all the data sets. 

\section{RESULTS AND DISCUSSIONS}
In Paper I, we studied the correlation between the EW of the \fekalfa line
 and the 2--10 keV continuum luminosity ($L_{\rm x}$) for the measurements from individual observations (``$per~observation$'' sample), 
and from individual sources (``$per~source$'' sample)\footnote{Measurements for
the latter were derived from only one spectrum per source,
which in some cases was the time-averaged spectrum of multiple observations, 
as described in Paper I.}, respectively. 
Despite better isolation of the \fekalfa line core,
there is still an anti-correlation between the EW and the X-ray luminosity in 
the diagrams (Figure 7 in Paper I). 
In this paper, we first present a similar analysis of the anti-correlation, this time 
including 15 new \chandra HEG observations. 
We found a similar result that 
 the ``$per~observation$'' sample gave a stronger anti-correlation ($\alpha=-0.18\pm0.03$) 
than the ``$per~source$'' sample ($\alpha=-0.11\pm0.03$). 
Quantitatively, the anti-correlation for the``$per~observation$'' results 
is significant at a level of $\sim6.39\sigma$, 
as opposed to $\sim3.74\sigma$ for the ``$per~source$'' results for a non-zero slope. 

\subsection{Fe K$\alpha$ Line Baldwin Effect in Individual Sources}
 
As we discussed in Paper I, the weaker anti-correlation (in the slope) in the ``$per~source$'' sample could be due to the obvious reduction of the EW variability when 
data from multiple HEG observations were averaged.
This clearly indicates that rapid variation in
X-ray continuum plays an important role in producing the XBE, and such
an effect should also be visible in individual AGNs.

In fact, the \fekalfa line EW in an individual source can vary by more 
than a factor of two.
For example, from Table 1 in Paper I, we can see that in five observations of NGC 4151, 
while the line intensity does not show 
significant variation, the continuum changes by a factor of $\sim$4 can lead to 
a change in the EW by a factor of $\sim$3-4. 
In Figure 1, we plot the 13 \chandra \hetg observations of NGC 4051 and the best-fit 
line for the correlation. We find clearly a decrease in the EW as the luminosity increases, 
an inverse correlation with $EW~\propto~L^{-0.75\pm0.60}$. 
{ Note that the slope cannot be well determined (the significance for the anti-correlation is only 
at a level of $1.24\sigma$), }
possibly due to the limited number of exposures and the large measurement errors
in Fe K$\alpha$ line EW. 
 
To make full use of multiple Chandra HEG observations on individual AGNs,
we plot in Figure 2 the EW of the \fekalfa line against 
the 2-10 keV luminosity for all sources which were observed by \chandra HEG more than once. Both the EW and the X-ray luminosity were normalized by the time-averaged values for each source. 
The averaged EW ($\langle \rm EW\rangle$) and X-ray luminosity ($\langle \rm L_x\rangle$) 
are the best-fitting values from the spectrum averaged over multiple observations 
of a source where relevant.  
It can be seen that there is clearly an inverse correlation between the normalized EW and 
luminosity in log-log space, thanks to our large sample of the HEG observations. 
Using a similar fitting method as in Paper I, 
we find the anti-correlation is significant at a level of $\sim7.8\sigma$, and 
the best-fitting slope of the correlation is --0.82$\pm0.10$, 
a value very close to --1 within the uncertainties. 
A slope of --1 could be expected if the \fekalfa flux in an individual AGN 
remains constant, since the EW was calculated by the flux of the 
\fekalfa line over the continuum at $\sim$ 6.4 keV. 
Note the slight deviation of the best-fit slope from --1 (see Figure 2) 
could be due to the less variable EW of the \fekalfa line than the 2-10 keV continuum. 
This is possibly because of spectral variability in some AGNs where the X-ray spectra soften as 
they brighten (with the relatively softer energies  
displaying stronger variability, see e.g., Markowitz \& Edelson 2004), 
so that the continuum variation amplitude at $\sim$ 6.4 keV could be smaller than 
that in the 2-10 keV. 
{
On the other hand, if there is an underlying broad Fe K$\alpha$
line component, it may affect the EW of the narrow line. However,
a broad-line component is difficult to deconvolve with the HEG. 
This problem was addressed in Yaqoob et al. (2001) who found that the
effect of an underlying broad line on the EW of the narrow line, with
parameters that are typical of type 1, is of the order of a few percent.
Since the narrow line dominates over the broad line even more in type 2
AGN, the effect is even less significant.
 
}

{

Another factor affecting the anti-correlation above could be the uncertainty in the X-ray 
continuum determination. 
The 2-10 keV continuum luminosities shown in Figure 2 were obtained by extrapolating 
the best-fitting model up to 10 keV, which could give inaccurate luminosities if the 
continuum shape is significantly different in the 
7-10 keV band compared to the extrapolated model. 
We performed simulations of 0.5--10 keV HEG spectrum based on the 
NGC 3783 data (the deepest observation with HEG) with a complex model 
including a partially covered 
absorption, a thermal black body for the soft excess, a power law and a 
Compton reflection component and both broad and narrow \fekalfa lines\footnote{In 
this model, we assumed a most complex continuum for AGNs in current sample, with absorption $N_H=10^{23}$ cm$^{-2}$ and 
Compton reflection component $R=4$. The broad Fe K Gaussian line was assumed to have 
$E=5.9$ keV, $\sigma=0.7$ keV and $EW=200$ eV (the values for MCG -6-30-15, e.g., Nandra et al. 2007).}. 
Using the simulated spectrum as input, we 
then fitted the 2-7 keV HEG data with the empirical power-law model as in Paper I and 
obtained the extrapolated 2-10 keV flux. We found a value is comparable with the input one, 
with a difference in the 2-10 keV X-ray flux of the order of 6\%. 
In addition, we also performed spectral fitting to the broad-band $Suzaku$ 
spectrum (0.5--50 keV) for one of our objects (F9), and compared the best-fitted 2-10 keV flux to 
that by extrapolating with the empirical model. We found that the difference in flux is 
less than $1$\%.  
We therefore conclude that 
the effect of the continuum slope at $\sim$ 7--10 keV on the values of 
the extrapolated X-ray luminosities is unimportant. 


}
\subsection{Simulating the Global X-ray Baldwin Effect}

Assuming constant \fekalfa line flux, Jiang et al. (2006) studied that whether the variability of the X-ray continuum alone can account for the global X-ray Baldwin
effect observed in their Composite Chandra/XMM sample.
Although having a large scatter, they found that the simulated anti-correlation has a slope of $-0.05\pm0.05$,
and 8.4\% of the simulations could produce anti-correlation steeper than the observed ones.
We checked this possibility for our \chandra grating sample
by simulating the variability of the EW that is caused by the X-ray continuum variations. 
The amplitude of the X-ray continuum variation can be quantified by calculating 
the fractional variability amplitude, $F_{\rm var}$ (e.g., Nandra et al.
1997a), which is simply the square root of the normalized excess variance ($\sigma^2_{\rm NXS}$), taking into account Poisson noise. 

In our simulations, we first adopted a similar time variation model as Jiang et al. (2006), 
that the observed X-ray luminosity is normally distributed 
with the Gaussian width being the fractional variability amplitude, 
using the correlation $F_{\rm var}\propto L_{\rm X}^{-0.135}$ of Markowitz \& Edelson (2004) 
for the longest timescale data (1296 days, $\sim$3.5 yr).  
Note that the use of the timescale of years is motivated by the fact 
that it corresponds to the 
expected light-crossing time of the region producing the narrow Fe~K line emission 
(see Shu et al. 2011). 

For comparison, we also computed the observed excess variance for our sources with
multiple HEG observations, 
based on the data from Figure 2. 
All sources were assumed to have an uncertainty in the continuum luminosity measurement, typically 
$\sim$2\% at 68\% confidence level. 
We obtained $\sigma^2_{\rm NXS}=0.054\pm0.012$, corresponding to a fractional variability 
amplitude $F_{\rm var}$ = 23.3\%. This is comparable with but smaller than the mean $F_{\rm var}$ of 33.3\% calculated from the employed time variation model, likely due to the fact that many of the multiple HEG observations spans a much shorter time scale.

To build mock samples with line EW that does not correlate with the X-ray 
luminosity, we first assign a constant line EW to all AGNs in our sample,
and apply a Gaussian scatter to the EW for each source (in log space, with 
$\sigma_{\rm rms}$EW measured by normalizing the observed EW of the 
''$per~source$'' results to the best-fitting line, see Fig. 7(c) in Paper I), to account for the
intrinsic scatter in EW distribution. 
Random continuum variations were then added to the luminosity for each source. 
To match the observed ''$per~source$'' sample in Paper I, multiple observations of 
individual sources were simulated and their luminosities were then averaged. 
Such simulation was repeated for each source 
to build 10,000 artificial samples.
Figure 3(a) shows the distribution of the best-fit slopes (solid line) for the artificial samples. 
It can be seen that $\sim$8\% simulations can produce steeper anti-correlation than that 
was observed ($\alpha=-0.11\pm0.03$),   
consistent with the results of Jiang et al. (2006).  

Recent observations of some highly variable narrow line Seyfert 1 galaxies and 
X-ray binaries revealed that some of them show lognormal X-ray flux distribution 
(e.g., Gaskell et al. 2004; Uttley et al. 2005). 
To better match the observations, it is worth considering the model of lognormal 
variability in our simulations, in addition to the above Gaussian luminosity distribution.
Because the $F_{\rm var}$ was calculated in linear space, the Gaussian width in 
log space was replaced by the formula $\sigma=\sqrt{ln(1+F_{\rm var}^2/2)}$.
We again built 10,000 artificial samples, and
the resulting distribution of the best-fit slopes is shown by the
 dot-dashed line in Figure 3 (a). 
It can be seen that, when using the lognormal X-ray continuum variations, the distribution of the slopes from the simulated datasets moves slightly rightwards, 
with $\sim$3\% of the simulations being able to produce 
the anti-correlation slopes steeper than the observed value. 

The lower panel in Figure 3 shows the distribution of the slopes from the simulations 
of the $"per~observation"$ sample. 
The solid and dot-dashed lines represent the distribution of the 
simulated anti-correlation slopes with
 the Gaussian and lognormal continuum variation models, respectively. 
We see results consistent with that of the ``$per~source$'' simulations,
that $\sim$11\% of the simulations with Gaussian variation model
can produce slopes steeper than the observed value ($\alpha=-0.18\pm0.03$), while
the lognormal model can only yield a fraction of $\sim$3\%.


As noted by Markowitz \& Edelson (2004), limitations exist 
when one uses the excess variance 
as a description of the intrinsic X-ray variability, 
since each light curve contains independent underlying 
stochastic process, and there would be significant random fluctuations in the 
measured variance. 
In addition, they found the strength of the X-ray variability tends to be different on 
different timescales and/or at different energy bands. 
With these caveats in the mind, we investigated the effect of different 
variability amplitudes on the simulations. 
Figure 4 shows distributions of the best-fit slopes of the simulated $"per~source"$ samples, 
with different variability amplitudes employed. 
It can be seen that the distribution of the simulated slopes moves towards  
much steeper values with increasing variability amplitudes. 
If the variability amplitude is increased by 50\%, 
$\sim$27\% of the simulations (Gaussian variation model) could produce anti-correlation slopes greater than the observed X-ray Baldwin effect, and the mean value is $0.07\pm0.04$, comparable with 
the observed value of $0.11\pm0.03$ within uncertainties.
In the case of lognormal variation model,  the fraction also increases from
3\% to 9\%.

\section{CONCLUSIONS}
In this paper, we presented a systematic study on the narrow \fekalfa line Baldwin effect in AGNs 
observed with the \chandra high energy grating. 
By normalizing the line EWs and continuum luminosities to the time-averaged values
for AGNs with multiple HEG observations, 
we found a strong anti-correlation between both quantities   
with a best-fitting slope of $-0.82\pm0.10$, statistically consistent with a slope of $-1$. 
This result indicates generally that the flux of the \fekalfa line of 
an individual AGN in \chandra grating sample is nearly constant and lacks 
 a corresponding response to the continuum variation.
We simulated the X-ray continuum variation 
to test the possibility whether such observational bias could totally account for the observed X-ray Baldwin effect. 
We find that the simulations could yield significant anti-correlation between EW and continuum luminosity, and the strength of the simulated Baldwin effect is sensitive
to the X-ray variation model and variation amplitude. 
Based on our HEG sample and current knowledge of AGN's X-ray variation, the possibility that the observed X-ray Baldwin effect is totally an observational bias due to X-ray variation cannot be ruled out.
This bias should be taken into account before any scientific conclusion can be made
based on an observed X-ray Baldwin effect.
One should endeavor to make as many repeated observations of each AGN as possible, 
covering the relevant timescales, and averaging the results. This 
could minimize the bias and help us to reveal
the underlying intrinsic X-ray Baldwin effect, if there is any.



X.W.S. thanks the support from China postdoctoral foundation. 
We acknowledge support from Chinese National Science Foundation 
(Grant No. 10825312, 11103017), and the Fundamental Research Funds for the Central Universities 
(Grant No. WK2030220004, WK2030220005). This research
made use of the HEASARC online data archive services, supported
by NASA/GSFC. 
The authors are grateful to the \chandra 
instrument and operations teams for making these observations
possible.

\begin{figure}
\epsscale{1.0}
\plotone{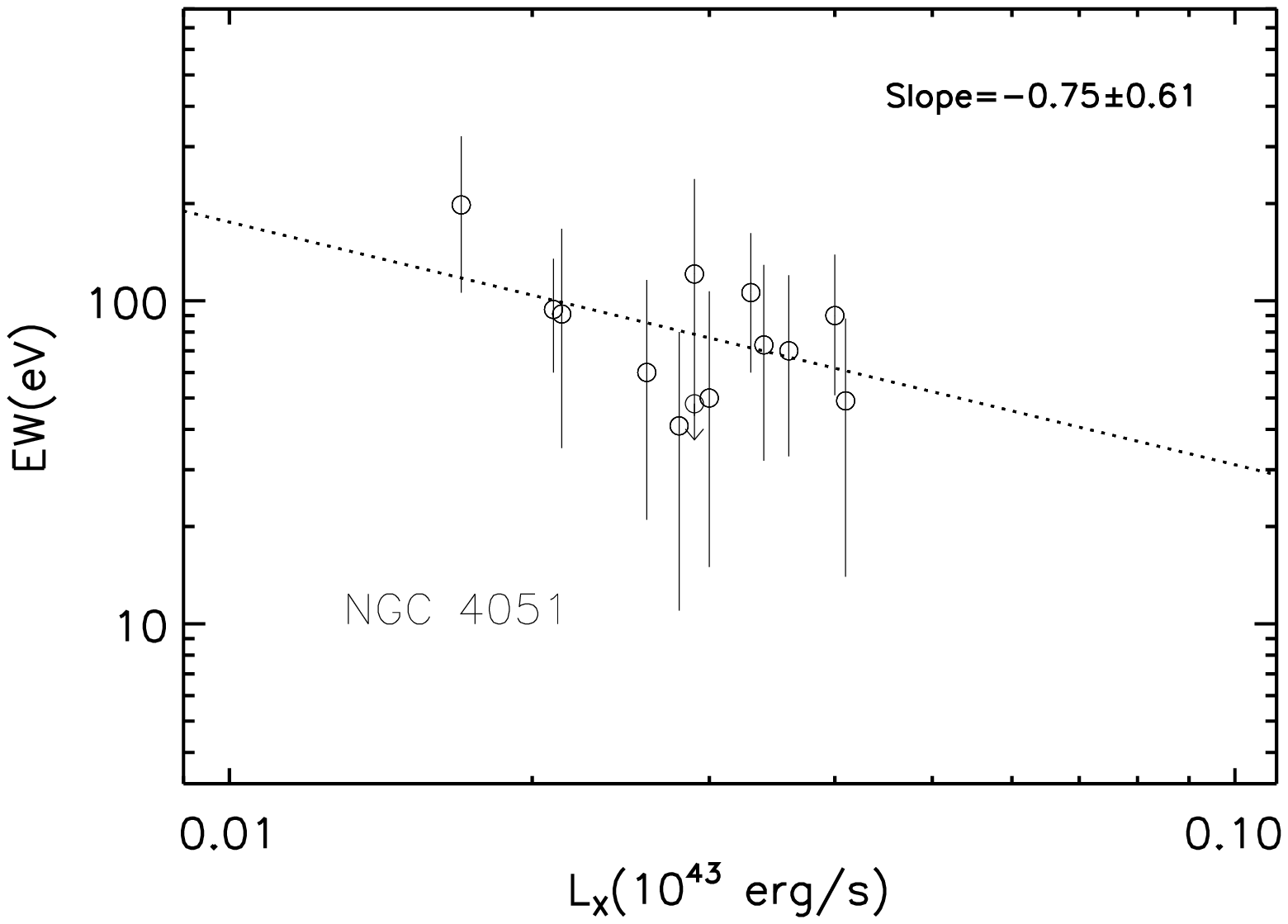}
\caption{
15 \chandra \hetg observations of NGC 4051, which show a clear
inverse correlation between the narrow \fekalfa line EW and the 
2--10 keV luminosity, with a very steep relationship $EW \propto L_x^{-0.75\pm0.61} $
(dotted line).}
 \end{figure}

\begin{figure}
\epsscale{1.0}
\plotone{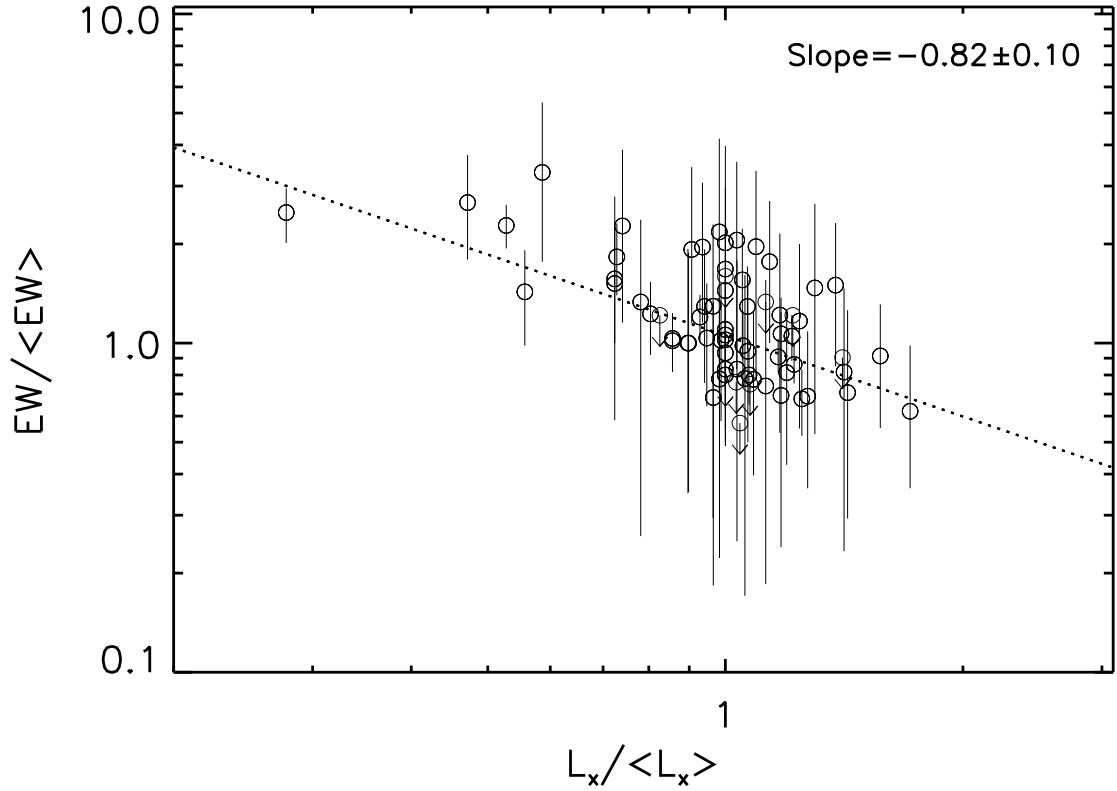}
\caption{
Relation between the \fekalfa line EW and the X-ray luminosity, both normalized by their
time-averaged values for sources which were observed more than once by \chandra HEG.
The time-averaged EW ($\langle EW \rangle$) and X-ray luminosity ($\langle L_x \rangle$) are
from the spectral fitting of the time-averaged spectrum for each source (see Paper I for details).
The dotted line shows the best fit to the data: $EW \propto L_x^{-0.82\pm0.10} $.
}
 \end{figure}

\begin{figure}
\epsscale{1.1}
\plotone{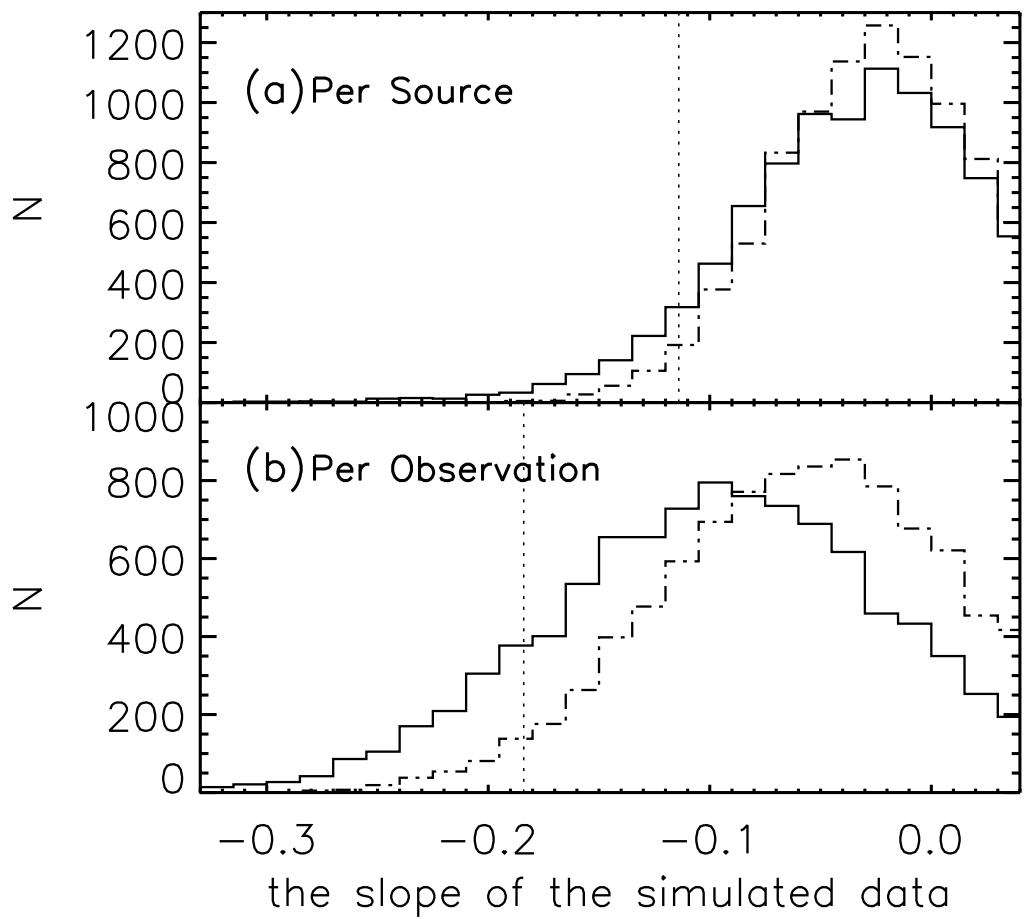}
\caption{
Distribution of the best-fit slopes of $\log EW$ versus $\log L_x$ of the simulated
datasets. {\it Panel (a)}: histogram of the simulated power-law slopes for ''$per~source$''
results. {\it Panel (b)}: as (a), but for ''$per~observation$'' sample.
The solid lines represent distributions assuming the variability is normally
(or Gaussian) distributed in the
simulations, while
dot-dashed lines are for the variability with lognormal distribution.
The vertical dotted lines show the observed slopes for each sample,
which are $-0.11\pm0.03$ and $-0.18\pm0.03$, respectively. }
\end{figure}

\begin{figure}
\epsscale{1.1}
\plotone{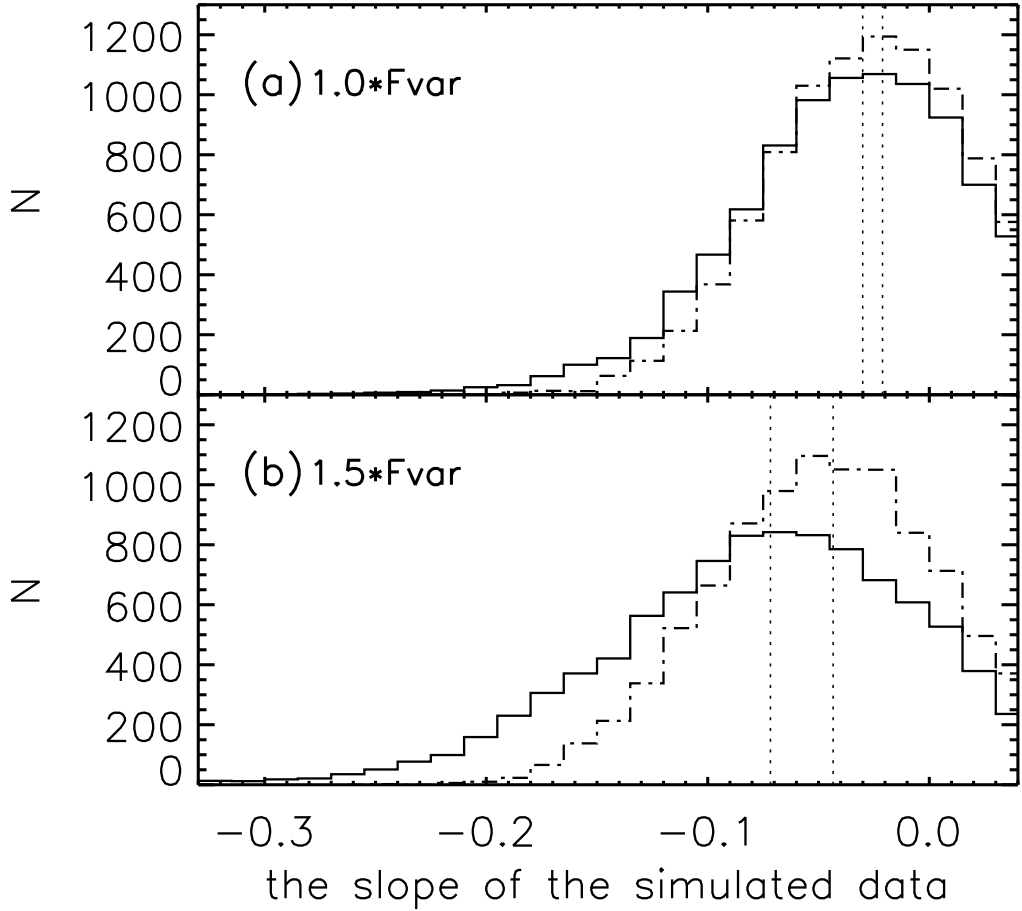}
\caption{
Distribution of the best-fit slopes of $\log EW$ versus $\log L_x$ for the
simulated ''$per~source$'' sample. The lower panel corresponds
to an increase of fractional variability amplitude ($F_{var}$) by 50\%.
The solid and dot-dashed lines represent different time variation model,
as shown in Figure 3.
The vertical dotted lines show the means of the slope for each distribution.
}\end{figure}

\end{document}